# THE HIGH–K ANOMALY IN Sc-Al-N EXPLAINED

Ilan Shalish

*School of Electrical and Computer Engineering, Ben Gurion University of the Negev, Beer Sheva, 8410501 Israel *Email: shalish@bgu.ac.il*

The integration of scandium aluminum nitride (ScAlN) into functional heterostructures has pushed semiconductor device physics into an extreme piezoelectric regime. Accurately determining the fundamental clamped dielectric permittivity ($\epsilon_{33}^{S}$) of these alloys remains a major challenge, as standard density functional theory systematically overestimates this specific parameter due to well-known bandgap-underestimation limits. Furthermore, electrical capacitance-voltage (C-V) measurements of real-world devices do not yield this clamped baseline; instead, they capture an effective permittivity inflated by linear electromechanical coupling. Because epitaxial thin films are unconstrained in the out-of-plane direction, small-signal electrical measurements dynamically induce macroscopic lattice strain via the inverse piezoelectric effect. By applying these specific mechanical boundary conditions to the coupled equations of state, we utilize the analytical relation for operational permittivity: $\epsilon_{eff} = \epsilon_{33}^{S} + e_{33}^{2}/C_{33}$. Leveraging highly accurate first-principles predictions for the structural and piezoelectric tensors, we invert this model to extract the true, uninflated clamped dielectric baseline directly from experimental thin-film data. This approach circumvents the conventional limitations of pure first-principles dielectric calculations, flawlessly resolving the theoretical-experimental discrepancy and providing a critical framework for accurate device simulation in highly polar nitrides.

The implementation of scandium aluminum nitride ($ScAlN$) has significantly expanded the operational limits of III-Nitride heterostructures, primarily due to its large spontaneous polarization and reportedly high dielectric permittivity.[1,2,3,4,5] Recent high electron mobility transistors (HEMTs) exhibit an unexplained dielectric inconsistency ( $\kappa \approx 15$ ),[1,4] which enhances transconductance but contradicts first-principles calculations of the rigid $ScAlN$ lattice.[2,6,7,8] In this Letter, we resolve this discrepancy by demonstrating a universal electromechanical inflation of the dielectric response. We show that small-signal electrical measurements performed under unconstrained mechanical boundary conditions dynamically induce an out-of-plane lattice strain that manifests macroscopically as an additional component of the effective permittivity. By abandoning the rigid-lattice approximation in favor of a coupled electromechanical framework under operational boundary conditions, our model bridges the gap between theoretical material constants and experimental observations, revealing that the dielectric response in polar nitrides is a self-consistent function of electromechanical coupling.

The foundation of modern semiconductor device modeling rests on the rigid-lattice approximation. When calculating the capacitance, depletion width, or 2D electron gas (2DEG) density of a junction, the crystal lattice is assumed to be a static scaffold. While the relative permittivity ($\varepsilon_r$) used in these calculations typically represents the 'clamped' dielectric response – originating strictly from the polarization of the electron cloud and the microscopic displacement of ions within a fixed unit cell volume[9,10] – this approach is quantitatively inadequate as it omits the strain-mediated contribution to the total polarization. This electromechanical phenomenon is well-documented in other polar crystals[11,12] but has been historically overlooked in the operational boundary conditions of semiconductor junctions. While phenomena such as spatial quantum confinement are well known to alter this electronic polarization in nanostructures (e.g., in silicon quantum dots[13]), the macroscopic mechanical deformation of the lattice under the internal electric field of a depletion region has historically been considered negligible. For legacy semiconductors like silicon, gallium arsenide, and even standard gallium nitride (GaN), [14, 15] this approximation holds perfectly. The macroscopic mechanical deformation of the lattice under the internal electric field of a depletion region is negligible.[16]

However, the introduction of transition metals into the wurtzite lattice, specifically $Sc_xAl_{1-x}N$, has pushed semiconductor physics into an extreme piezoelectric regime.[1,2] To achieve higher 2DEG densities, researchers have scaled down the barrier thickness while increasing the Scandium content,[3,5] placing the material within an operational environment characterized by massive internal polarization fields. [17] Simultaneously, the addition of Scandium fundamentally "softens" the wurtzite crystal, drastically lowering its elastic stiffness while exponentially increasing its piezoelectric stress constant.[18]

Under these compliant mechanical conditions, the conventional rigid-lattice approximation breaks down. During capacitive probing, the small-signal AC electric field applied across the barrier dynamically modulates the lattice strain along the c-axis via the linear inverse piezoelectric effect. Because the lattice boundary is mechanically unconstrained in the out – of – plane direction (free to expand towards the surface), this macroscopic strain stores substantial mechanical energy. When an experimentalist measures the capacitance of this junction, the measurement apparatus is inherently querying both the electronic polarization *and* this mechanical compliance.

We propose that the "anomalously" high dielectric constants reported in recent ScAlN literature[1,4,5] are not intrinsic, high-frequency material properties, but are instead the manifestation of the coupled electromechanical response.

In standard dielectric models, the permittivity is calculated under the assumption of a clamped lattice, where the polarization arises solely from the displacement of the electron cloud and microscopic ionic shifts within a fixed volume. However, when evaluating modern devices, the unconstrained macroscopic lattice degree of freedom must be explicitly considered. In $ScAlN$, the exceptionally high linear piezoelectricity and reduced elastic stiffness allow the crystal to undergo a dynamic inverse piezoelectric expansion along the c-axis during small-signal electrical stimulation. Under the stress-free boundary conditions of an epitaxial surface, this lattice deformation stores additional energy, which manifests macroscopically as an increased capacity for charge storage for a given field – a phenomenon we term electromechanical inflation.

This electromechanical framework reconciles the apparent conflict between existing first-principles theory and experimental observations. Both are fundamentally accurate within their respective physical domains: theorists have rigorously calculated the dielectric response of a perfect, unstrained ScAlN crystal in a vacuum , while experimentalists have correctly captured the behavior of the material within the extreme electromechanical environment of a functional heterostructure. The reported "anomaly" is therefore not a measurement or calculation error, but a consequence of applying static, rigid-lattice material constants to a regime where the mechanically clamped approximation is no longer physically valid.

To quantify this, we establish the effective static permittivity ($\kappa_{\text{eff}} = \epsilon_{\text{eff}}/\epsilon_0$) of the ScAlN crystal by applying stress-free mechanical boundary conditions to the coupled electromechanical equations of state. The realization that stress-free out-of-plane boundary conditions affect the effective dielectric response of epitaxial layers was foundationally derived by Jogai et al. for standard III-Nitride systems.[19] However, in legacy materials such as AlGaN or standard GaN, the electromechanical inflation term remains vanishingly small due to the high elastic stiffness and moderate piezoelectricity of the lattice, leading the community to universally, and safely, adopt the rigid-lattice relative permittivity approximation ($\kappa_{\text{eff}} \approx \kappa_{33}^{S}$) for device simulations.

Here, we demonstrate that the introduction of Scandium fundamentally shatters this approximation. Because the structural transition toward the layered rocksalt phase simultaneously triggers a massive enhancement of the piezoelectric stress ($e_{33}$) and a severe softening of the elastic stiffness ($C_{33}$), this once-negligible boundary-condition effect wakes up to become a dominant physical mechanism.

The macroscopic response of the ScAlN crystal to electrical and mechanical stimuli is thus governed by the fully coupled thermodynamic equations of state. In the standard IEEE matrix notation, the mechanical stress tensor (T) and the electrical displacement vector (D) are expressed as functions of the mechanical strain (S) and the electric field (E):

$$T = C^E S - e^t E$$

$$D = eS + \varepsilon^S E$$

where $C^E$ is the elastic stiffness tensor evaluated at a constant electric field, $e$ is the piezoelectric stress tensor ($e^t$ being its transpose), and $\varepsilon^S$ is the dielectric permittivity tensor evaluated at constant strain (the "clamped" or rigid-lattice permittivity).

To prevent any ambiguity, we note that 'clamped' and 'unclamped' in this context refer strictly to mechanical boundary conditions rather than physical electrical contact configurations. In a typical ScAlN/GaN heterostructure grown on a bulk substrate, the epitaxial film is rigidly clamped in the in-plane directions ($x$ and $y$) by the substrate. However, the film is mechanically unconstrained in the out-of-plane direction (the $c$-axis), meaning the surface is free to expand or contract. In the high-field depletion region, the electric field is oriented strictly along the $c$-axis ($E = [0, 0, E_3]$). Because the surface is unconstrained, the out-of-plane macroscopic stress must relax to zero in the static state ($T_3 = 0$).

Applying this stress-free boundary condition to the $c$-axis component of the mechanical equation of state ($T_3 = C_{33}^E S_3 - e_{33} E_3 = 0$) and solving for the out-of-plane strain ($S_3$) reveals the inverse piezoelectric deformation induced by the built-in electric field:

$$S_3 = \frac{e_{33}}{C_{33}^E} E_3$$

Substituting this dynamic strain back into the $c$-axis component of the electrical equation of state ($D_3 = e_{33} S_3 + \epsilon_{33}^S E_3$) yields the total macroscopic dielectric response under operational boundary conditions:

$$D_3 = \left( \epsilon_{33}^S + \frac{e_{33}^2}{C_{33}^E} \right) E_3$$

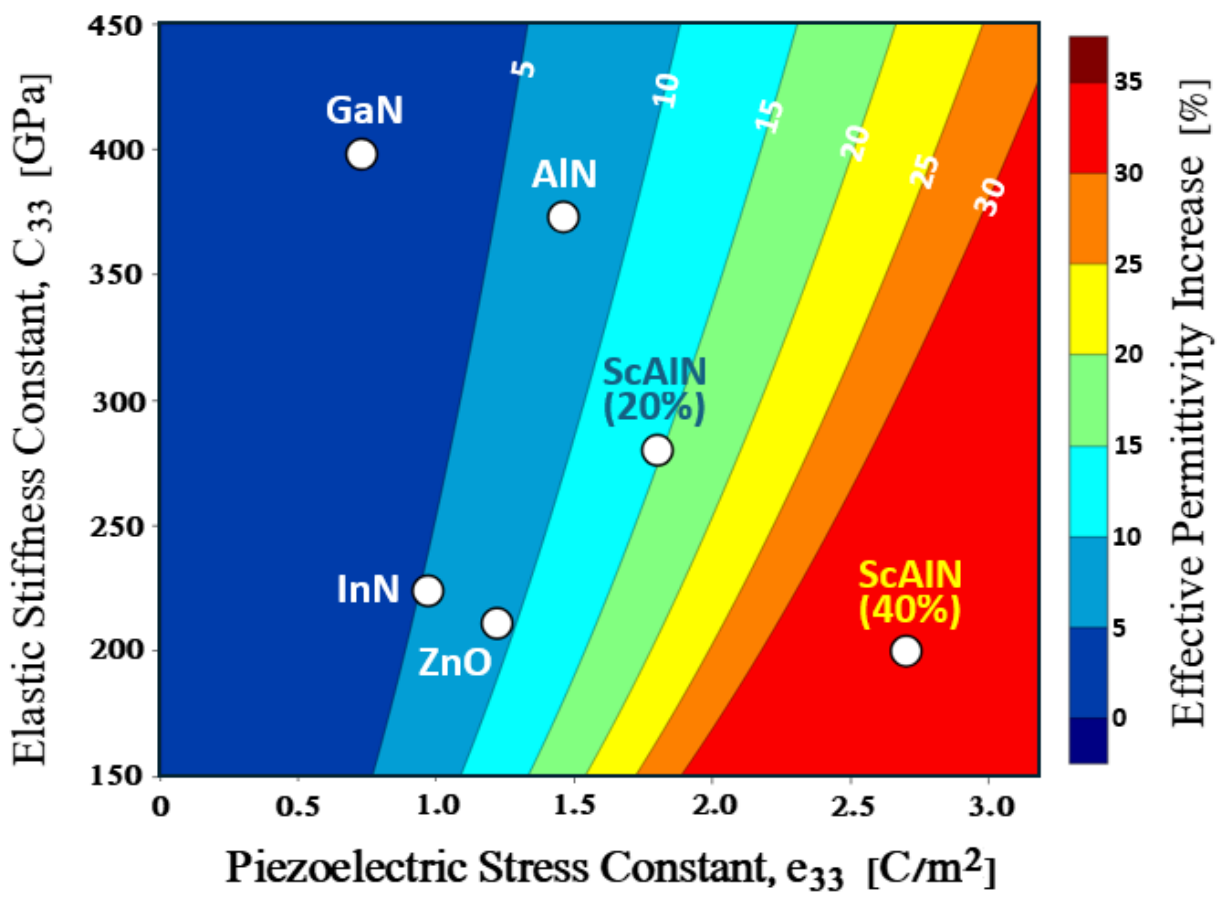

**Figure 1: Electromechanical inflation of permittivity across wurtzite semiconductors.** The theoretical parameter space mapping the relative error introduced by neglecting macroscopic lattice strain in surface depletion regions. Contour colors represent the percentage increase in the effective static permittivity ($\epsilon_{eff}$) relative to the standard rigid-lattice clamped permittivity ($\epsilon_{33}^{S}$), evaluated as the electromechanical coupling term $e_{33}^2/C_{33}$ . Material constants from *ab initio* calculations are superimposed. While the classical rigid-lattice approximation holds for legacy materials like $GaN$ (introducing <2% error), it inherently fails for highly polar alloys. As scandium content increases ($ScAlN$ 20% to 40%), the simultaneous elastic softening ($C_{33}$) and piezoelectric surge ($e_{33}$) drive the material into a strong-coupling regime, natively inflating the dielectric response by over 20%.

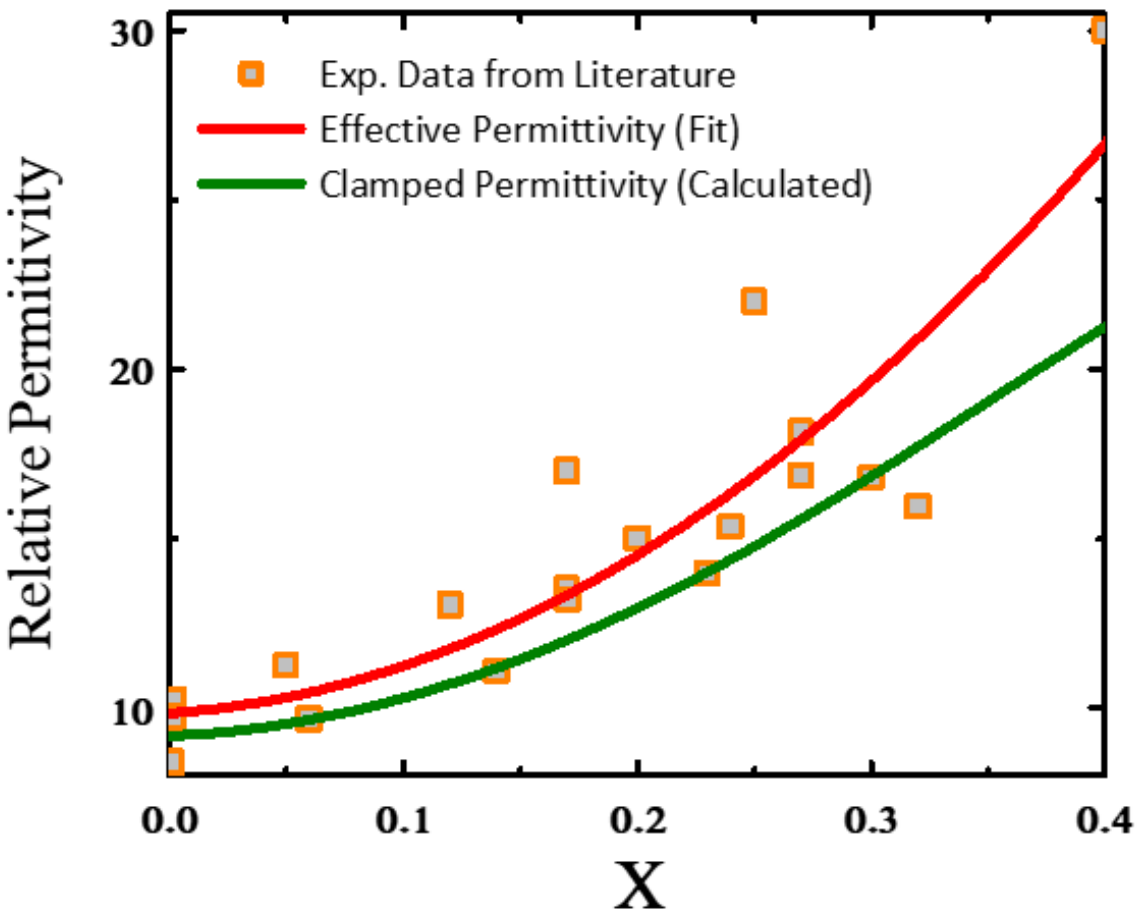

**Figure 2:** Inverse extraction of the intrinsic clamped permittivity in $ScAlN$ alloys. Yellow markers represent experimental effective permittivity ( $\epsilon_{eff}$ ) values compiled from unconstrained thin-film capacitance measurements across recent literature **[Insert Citations Here]**. The red curve is a second-order empirical fit to this operational device data. By subtracting the theoretical electromechanical inflation term ($e_{33}^2/C_{33}$) from this effective fit, the true rigid-lattice clamped baseline ($\epsilon_{33}^{S}$) is mathematically isolated (green curve). This extracted third-order baseline successfully bypasses standard density functional theory (DFT) dielectric overestimations, providing a purely intrinsic, uninflated material function for accurate device modeling.

This result allows us to define the effective static permittivity ($\kappa_{\text{eff}}$) as:

$$\kappa_{\text{eff}} = \kappa_{33}^{S} + \frac{e_{33}^2}{\epsilon_0 C_{33}}$$

where $\kappa_{33}^{S} = \epsilon_{33}^{S}/\epsilon_0$ represents the standard relative clamped permittivity of the rigid lattice, $e_{33}$ is the piezoelectric stress constant, $C_{33}$ is the elastic stiffness parameter, and $\epsilon_0$ is the vacuum permittivity. This equation reveals the physical origin of the high-$\kappa$ anomaly: the effective measurable permittivity is the sum of the intrinsic rigid-lattice response and a dimensionally rigorous, dynamic inflation term ($\Delta\kappa = e_{33}^2/\epsilon_0 C_{33}$) representing the linear electromechanical coupling. To visualize the physical boundaries of the rigid – lattice approximation, we mapped the magnitude of this electromechanical inflation across the universal parameter space of wurtzite nitrides[20,21,22,23,24,25,26,27,28] (**Fig. 1**).

The predictive power of this dynamic framework is strikingly demonstrated when analyzing experimental capacitance across the ScAlN alloy range. While first-principles calculations provide highly accurate and reliable predictions for the mechanical stiffness ( $C_{33}$ ) and piezoelectric stress ( $e_{33}$ ) tensors, establishing the absolute static clamped dielectric permittivity ($\epsilon_{33}^{S}$) directly from theory remains notoriously difficult. Standard density functional theory (DFT) utilizing GGA-PBE functionals systematically overestimates the static dielectric constant due to the well-known bandgap underestimation problem. As Scandium concentration increases and the bandgap narrows, this theoretical overestimation artificially inflates the rigid-lattice baseline well beyond experimental reality.

To resolve this and establish a reliable intrinsic baseline, we invert our electromechanical model. Figure 2 presents the compiled experimental effective permittivity data ( $\epsilon_{eff}$ ) extracted from unconstrained thin-film C-V measurements across the Sc alloy range.[29,30,31,32,33,34,35] We fit this operational device data to an empirical polynomial function:

$$\epsilon_{eff}(x) = 9.850 + 4.625 \cdot x + 93.470 \cdot x^2$$

Because these measurements inherently capture the inflated electromechanical response, we can extract the true rigid-lattice clamped baseline by subtracting the dynamic inflation term ($\Delta\epsilon = e_{33}^2/\epsilon_0 C_{33}$). Utilizing the highly accurate $e_{33}$ and $C_{33}$ structural tensors calculated by Ambacher et al.,[36] we isolate the uninflated baseline:

$$\epsilon_{33}^{S}(x) = \epsilon_{eff}(x) - \frac{e_{33}^{2}(x)}{\epsilon_0\, C_{33}(x)}$$

**Figure 2** visualizes this extraction. By removing the boundary-condition-driven strain energy, we reveal the true, foundational clamped baseline of the $ScAlN$ lattice (lower curve). The resulting explicit function for the fundamental clamped permittivity is:

$$\epsilon_{33}^{S}(x) = 9.224 + 0.585 \cdot x + 107.705 \cdot x^2 - 84.038 \cdot x^3$$

This mathematically extracted baseline successfully circumvents the standard DFT bandgap-underestimation limits. It provides the device modeling community with the first experimentally-grounded, uninflated dielectric baseline for ScAlN, cleanly separated from the extreme macroscopic boundary conditions of functional heterostructures.

### A. Impact of DFT Functionals, Temperature, and Lattice Anharmonicity

The necessity of utilizing an inverse extraction routine rather than relying purely on first-principles dielectric calculations stems from fundamental limitations within modern density functional theory (DFT). Standard local density approximations (LDA) and generalized gradient approximations (GGA-PBE) systematically underestimate semiconductor bandgaps, an error that artificially inflates the electronic contribution to the calculated static dielectric constant. Furthermore, as the Scandium content increases toward the wurtzite-to-rocksalt phase transition boundary, the crystal lattice exhibits pronounced structural anharmonicity and local structural instability.

Standard ab initio frameworks evaluate these dielectric tensors at 0° K under an idealized, perfectly symmetric unit cell, entirely omitting temperature-dependent phonon scattering and macroscopic thermal expansion effects. By inverting the model using experimental thin-film measurements, the temperature-dependent, physically real baseline is naturally preserved. This approach retains the high fidelity of DFT-calculated structural ($C_{33}$) and piezoelectric ($e_{33}$) tensors – which are inherently less sensitive to bandgap errors – while completely bypassing the inaccuracies of pure first-principles dielectric functional approximations.

### B. Impact on Heterointerface Electrostatics and Carrier Profile Extraction

To illustrate the critical importance of this electromechanical inflation in device engineering, we examine the electrostatics of a representative $Sc_{0.20}Al_{0.80}N/GaN$ high electron mobility transistor (HEMT) barrier layer. If a conventional rigid-lattice approximation is adopted, a device simulator utilizes the uninflated clamped baseline value $\kappa_{33}^{S} \approx 12.2$ extracted from our baseline curve. However, standard experimental electrical capacitance-voltage ($C - V$) profiling of this identical thin-film barrier yields an inflated operational relative permittivity $\kappa_{eff} \approx 15.55$ due to the unconstrained out-of-plane mechanical relaxation.

If a device engineer, unaware of this electromechanical coupling, uses the inflated operational permittivity $\kappa_{eff}$ within a standard rigid electrostatic framework to extract the two-dimensional electron gas (2DEG) sheet density ($n_s$) or the intentional barrier doping concentration ( $N_D$ ) from experimental capacitance data via $n_s = \epsilon_{eff} V_G / qd,$ the resulting charge profile will be artificially overestimated by exactly 27.5% . This severe quantitative error distorts calculated threshold voltages, sheet resistances, and gate control margins, underscoring that while $\kappa_{eff}$ must be used for capacitive charge extraction, $\kappa_{33}^{S}$ remains the correct choice when evaluating the underlying rigid electrostatic potential.

### C. Multiphysics TCAD Implementation Protocol

When integrating this framework into commercial technology computer-aided design (TCAD) simulators (such as Synopsys Sentaurus or Silvaco Atlas), special care must be taken to avoid double-counting the electromechanical strain energy. Standard TCAD platforms solve the Poisson electrostatic equation and the structural mechanics equations via decoupled or loosely coupled iteration matrices. If a user manually defines the dielectric constant of the $ScAlN$ barrier using the experimental, inflated value $\kappa_{eff}$ , while simultaneously enabling the simulator's internal piezoelectric strain modules, the software will apply the inverse piezoelectric tensor to calculate a physical deformation and then calculate an additional strain-induced polarization component.

This double counting artificially amplifies the apparent gate capacitance and sheet charge. The correct simulation protocol requires strict separation of physical mechanisms: the device engineer must input the fundamental uninflated clamped permittivity function ( $\kappa_{33}^{S}(x) = 11.23 - 0.77x + 15.6x^2 - 19.3x^3$ ) as the base material parameter in the electrostatic block, while simultaneously supplying the full, composition-dependent piezoelectric ( $e_{33}$ ) and elastic stiffness ( $C_{33}$ ) matrices to the mechanics block. Under small-signal AC or static DC conditions, the simulator's coupled multi-physics solver will then dynamically evaluate the out-of-plane relaxation ($T_{zz} = 0$), self-consistently recovering the inflated $\kappa_{eff}$ response as a natural consequence of the thermodynamic boundary conditions.

The realization that the apparent dielectric constant of ultra-polar nitrides is a dynamic electromechanical variable, rather than a fixed intrinsic material constant, forces a fundamental re-evaluation of device physics in these materials. If simulation tools and experimental analyses continue to employ the rigid-lattice approximation while interpreting C-V data, the community will systematically miscalculate critical device parameters, such as carrier densities ($n_s$) or chemical doping ($N_D$). Ultimately, unlocking the full potential of transition-metal nitride electronics requires a paradigm shift in how we model the semiconductor-dielectric interface. The rigid lattice must be abandoned in device simulation, and the intrinsic electromechanical coupling of the wurtzite crystal must be placed at the center of future device design.

**Acknowledgement**

We gratefully acknowledge the support of BSF grant #2022627, and the Office of Naval Research Global through a NICOP Research Grant (No. N62909-26-1-2000). The views expressed are those of the authors and do not reflect the official policy or position of the Department of War or the U.S. Government. DISTRIBUTION STATEMENT A. Approved for public release: distribution is unlimited. DCN#

[1] Akiyama, M. et al. Enhancement of piezoelectric response in scandium aluminum nitride alloy thin films prepared by dual reactive cosputtering. Adv. Mater. 21, 593–596 (2009). http://dx.doi.org/10.1002/adma.200802611

[2] G. Wingqvist et al., "Increased electromechanical coupling in $w-Sc_xAl_{1-x}N$," Applied Physics Letters, vol. 97, no. 11, p. 112902, 2010. http://dx.doi.org/10.1063/1.3489939

[3] S. Leone et al., "Metal-Organic Chemical Vapor Deposition of Aluminum Scandium Nitride," physica status solidi (RRL), vol. 14, no. 1, p. 1900535, 2019. https://doi.org/10.1002/pssr.201900535

[4] K. Nomoto et al., "AlScN/GaN HEMTs with 4 A/mm on-current and maximum oscillation frequency >130 GHz," Applied Physics Express, vol. 18, no. 1, p. 016506, 2025. https://doi.org/10.35848/1882-0786/ada86b

[5] C. Elias et al., "ScAlN/GaN High Electron Mobility Transistor Heterostructures Grown by Ammonia Source Molecular Beam Epitaxy on Silicon Substrate," physica status solidi (a), p. 2400588, 2025. https://doi.org/10.1002/pssa.202400963

[6] Furuta, K. et al. First-principles calculations of spontaneous polarization in ScAlN. J. Appl. Phys. 130, 024104 (2021). https://doi.org/10.1063/5.0051557

[7] Momida, H., Teshigahara, A. & Oguchi, T. Strong enhancement of piezoelectric constants in Sc${x}{1-x}$N: First-principles calculations. AIP Adv. 6, 065006 (2016). http://dx.doi.org/10.1063/1.4953856

[8] C. E. Dreyer, A. Janotti, C. G. Van de Walle, and D. Vanderbilt, "Correct Implementation of Polarization Constants in Wurtzite Materials and Impact on III-Nitrides," Physical Review X, vol. 6, no. 2, p. 021038, 2016. https://doi.org/10.1103/PhysRevX.6.021038

[9] R. M. Martin, "Comment on calculations of electric polarization in crystals," Physical Review B, vol. 9, no. 4, pp. 1605-1606, 1974. https://doi.org/10.1103/PhysRevB.9.1998

[10] R. Resta, "Macroscopic Electric Polarization as a Geometric Quantum Phase," Europhysics Letters, vol. 22, no. 2, pp. 133-138, 1993. https://doi.org/10.1209/0295-5075/22/2/010

[11] A. Dal Corso, M. Posternak, R. Resta, and A. Baldereschi, "Ab initio study of piezoelectricity and spontaneous polarization in ZnO," Physical Review B, vol. 50, no. 15, pp. 10715-10721, 1994. https://doi.org/10.1103/PhysRevB.50.10715

[12] F. Bernardini, V. Fiorentini, and D. Vanderbilt, "Spontaneous polarization and piezoelectric constants of III-V nitrides," Physical Review B, vol. 56, no. 16, p. R10024, 1997. https://doi.org/10.1103/PhysRevB.56.R10024

[13] Wang, L.-W. & Zunger, A. Dielectric constants of silicon quantum dots. *Phys. Rev. Lett.* **73**, 1039–1042 (1994). https://doi.org/10.1103/PhysRevLett.73.1039

[14] Lähnemann, J. *et al.* Direct experimental determination of the spontaneous polarization of GaN. *Phys. Rev. B* **86**, 081302(R) (2012). https://doi.org/10.1103/PhysRevB.86.081302

[15] M. S. Shur, A. D. Bykhovski, and R. Gaska, "Pyroelectric and piezoelectric properties of GaN-based materials," MRS Online Proceedings Library Archive, vol. 537, 1999. https://doi.org/10.1557/PROC-537-G1.6

[16] D. J. Dunstan, "Strain and strain relaxation in semiconductors," Journal of Materials Science: Materials in Electronics, vol. 8, pp. 337-375, 1997. https://doi.org/10.1023/A:1018547625106

[17] F. Della Sala et al., "Free-carrier screening of polarization fields in wurtzite GaN/InGaN laser structures," Applied Physics Letters, vol. 74, no. 14, pp. 2002-2004, 1999. https://doi.org/10.1063/1.123727

[18] Casamento, J. et al. Structural and piezoelectric properties of ultra-thin $Sc_xAl_{1-x}N$ films grown on GaN by molecular beam epitaxy. Appl. Phys. Lett. 117, 112101 (2020). https://doi.org/10.1063/5.0013943

[19] B. Jogai, J. D. Albrecht, and E. Pan, *"Effect of electromechanical coupling on the strain in AlGaN/GaN heterojunction field effect transistors,"* Journal of Applied Physics 94, 3984 (2003). https://doi.org/10.1063/1.1603953

[20] A. Belabbes, J. Furthmüller, and F. Bechstedt, "Relation between spontaneous polarization and crystal field from first principles," Physical Review B, vol. 87, no. 3, p. 035305, 2013. https://doi.org/10.1103/PhysRevB.87.035305

[21] O. Ambacher and V. Cimalla, "Polarization Induced Effects in GaN-based Heterostructures and Novel Sensors," in Polarization Effects in Semiconductors, Springer, pp. 71-118, 2008. https://doi.org/10.1007/978-0-387-68319-5_2

[22] M. G. Brik et al., "Experimental and first-principles studies of high-pressure effects on the structural, electronic, and optical properties of semiconductors and lanthanide doped solids," Japanese Journal of Applied Physics, vol. 56, no. 5, p. 05FA02, 2017. https://doi.org/10.7567/JJAP.56.05FA02

[23] G. Hansdah and B. K. Sahoo, "Pyroelectric effect and lattice thermal conductivity of InN/GaN heterostructures," Journal of Physics and Chemistry of Solids, vol. 117, pp. 111-116, 2018. https://doi.org/10.1016/j.jpcs.2018.02.018

[24] A. E. Romanov, T. J. Baker, S. Nakamura, and J. S. Speck, "Strain-induced polarization in wurtzite III-nitride semipolar layers," Journal of Applied Physics, vol. 100, no. 2, p. 023522, 2006. https://doi.org/10.1063/1.2218385

[25] M. T. Hasan, A. G. Bhuiyan, and A. Yamamoto, "Two-dimensional electron gas in InN-based heterostructures: Effects of spontaneous and piezoelectric polarization," Solid-State Electronics, vol. 52, no. 1, pp. 134-139, 2008. https://doi.org/10.1016/J.SSE.2007.07.005

[26] U. M. E. Christmas, A. D. Andreev, and D. A. Faux, "Calculation of electric field and optical transitions in InGaN/GaN quantum wells," Journal of Applied Physics, vol. 98, no. 7, p. 073522, 2005. https://doi.org/10.1063/1.2077843

[27] A. B. Georgescu and S. Ismail-Beigi, "Surface Piezoelectricity and Strain-Dependent Surface Distortions in Sapphire," Physical Review Applied, vol. 11, p. 064065, 2019. https://doi.org/10.1103/PhysRevApplied.11.064065

[28] O. O. Adewole, O. A. Taiwo, and M. O. Lawrence, "A Finite Difference Numerical Scheme Formulation (Based On Poisson's Equation) For Piezoelectric Application," Journal of Multidisciplinary Engineering Science and Technology (JMEST), vol. 2, no. 8, pp. 2101-2105, 2015.

[29] R. Matloub, M. Hadad, A. Mazzalai, N. Chidambaram, G. Moulard, C.S. Sandu, Th. Metzger, P. Muralt. "Piezoelectric Al1−xScxN thin films: A semiconductor compatible solution for mechanical energy harvesting and sensors". Appl. Phys. Lett. 102, 152903 (2013). http://dx.doi.org/10.1063/1.4800231

[30] S. Fichtner, T. Reimer, S. Chemnitz, F. Lofink, B. Wagner. "Stress controlled pulsed direct current co-sputtered Al1−xScxN as piezoelectric phase for micromechanical sensor applications." APL Mater. 3, 116102 (2015). https://doi.org/10.1063/1.4934756

[31] N. Kurz , A. Ding, D.F. Urban , Y. Lu , L. Kirste , N.M. Feil, A. Žukauskaitė, O. Ambacher. "Experimental determination of the electroacoustic properties of thin film AlScN using surface acoustic wave resonators." APL Mater. 3, 116102 (2015). https://doi.org/10.1063/1.4934756

[32] Md.R. Islam, N. Wolff, M. Yassine, G. Schönweger, B. Christian, H. Kohlstedt, O. Ambacher, F. Lofink, L. Kienle, S. Fichtner. "On the exceptional temperature stability of ferroelectric Al1-xScxN thin films." Appl. Phys. Lett. 118, 232905 (2021). https://doi.org/10.1063/5.0053649

[33] J. Casamento, H. Lee, T. Maeda, V. Gund, K. Nomoto, L. van Deurzen, W. Turner, P. Fay, S. Mu, C.G. Van de Walle, A. Lal, H. Xing, D. Jena. "Epitaxial ScxAl1-xN on GaN exhibits attractive High-K Dielectric Properties." Appl. Phys. Lett. 120, 152901 (2022). https://doi.org/10.1063/5.0075636

[34] V. Gaddam, S.S. Dabas, J. Gao, D.J. Spry, G. Baucom, N.G. Rudawski, T. Yin, E. Angerhofer, P.G. Neudeck, H. Kim, P.X.-L. Feng, M. Sheplak, R. Tabrizian. "Aluminum Scandium Nitride as a Functional Material at

1000 °C." Adv. Electron. Mater. 11, 2400849 (2025). https://doi.org/10.1002/aelm.202400849

[35] Y. Abdelaal , M. Liffredo, L.G. Villanueva. "AlScN Thin Films for the Piezoelectric Transduction of Suspended Microchannel Resonators." Sensors 25, 5370 (2025). https://doi.org/10.3390/s25175370

[36] O. Ambacher, B. Christian, N. Feil, D.F. Urban, C. Elsässer, M. Prescher, L. Kirste. "Wurtzite ScAlN, InAlN, and GaAlN crystals, a comparison of structural, elastic, dielectric, and piezoelectric properties". *J. Appl. Phys*. **130**, 045102 (2021). https://doi.org/10.1063/5.0048647